\documentclass[preprint,12pt]{elsarticle}

\usepackage{amssymb}
\usepackage{amsmath}
\usepackage{amsfonts}
\usepackage{amsthm}
\usepackage{graphicx}

\DeclareMathOperator{\Hv}{\Theta}
\DeclareMathOperator{\Rp}{R}

\journal{JTB}

\begin{document}

\newcommand{\od}{\, \mathrm{d}}
\newcommand{\pd}{\partial}
\newcommand{\cc}{\mathrm{c.c.}}
\newcommand{\bs}[1]{\boldsymbol{#1}}
\newcommand{\lb}{\left}
\newcommand{\rb}{\right}
\newcommand{\fr}[2]{\frac{#1}{#2}}

\begin{frontmatter}



\title{Host control and nutrient trading in a photosynthetic symbiosis}

\author[YorkBio,YorkMaths]{Andrew Dean\corref{email}}
\author[YorkBio]{Ewan Minter}
\author[YorkBio]{Megan S\o renson}
\author[Exeter]{Christopher Lowe}
\author[Sheffield]{Duncan Cameron}
\author[YorkBio]{Michael Brockhurst}
\author[YorkBio,YorkMaths]{A. Jamie Wood}

\cortext[email]{andrew.dean@york.ac.uk}

\address[YorkBio]{Department of Biology, University of York, Wentworth Way, York YO10 5DD, UK}

\address[YorkMaths]{Department of Mathematics, University of York, York YO10 5DD, UK}

\address[Exeter]{Centre for Ecology and Conservation, College of Life and Environmental Sciences, University of Exeter, Penryn Campus, Treliever Road, Penryn TR10 9FE, UK}

\address[Sheffield]{Department of Animal and Plant Sciences, Alfred Denny Building, University of Sheffield, Western Bank, Sheffield S10 2TN, UK}


\begin{abstract}
Photosymbiosis is one of the most important evolutionary trajectories, resulting in the chloroplast and the subsequent development of all complex photosynthetic organisms. The ciliate \emph{Paramecium bursaria} and the alga \emph{Chlorella} have a well established and well studied light dependent endosymbiotic relationship. Despite its prominence there remain many unanswered questions regarding the exact mechanisms of the photosymbiosis. Of particular interest is how a host maintains and manages its symbiont load in response to the allocation of nutrients between itself and its symbionts. Here we construct a detailed mathematical model, parameterised from the literature, that explicitly incorporates nutrient trading within a deterministic model of both partners. The model demonstrates how the symbiotic relationship can manifest as parasitism of the host by the symbionts, mutualism, wherein both partners benefit, or exploitation of the symbionts by the hosts. We show that the precise nature of the photosymbiosis is determined by both environmental conditions (how much light is available for photosynthesis) and the level of control a host has over its symbiont load. Our model provides a framework within which it is possible to pose detailed questions regarding the evolutionary behaviour of this important example of an established light dependent endosymbiosis; we focus on one question in particular, namely the evolution of host control, and show using an adaptive dynamics approach that a moderate level of host control may evolve provided the associated costs are not prohibitive.
\end{abstract}


\end{frontmatter}


\section{Introduction}

Endosymbiotic relationships are widespread in nature and play key roles in the functioning of many ecosystems \cite{esteban2010mixotrophy, howells2012coral, jeong2010growth, johnson2011acquired, johnson2011acquisition, moorthi2009mixotrophy, stoecker2009acquired}. Different symbioses have evolved many times throughout history; of particular note is the evolution of cellular organelles such as chloroplasts from a cyanobacteria-eukaryote symbiosis \cite{keeling2013number}. One well-known example of endosymbiosis is the relationship between the ciliate \emph{Paramecium bursaria} and the alga \emph{Chlorella} \cite{karakashian1963growth}. These organisms \cite{karakashian1963growth, fujishima2009endosymbionts} and their close relatives \cite{esteban2010mixotrophy, fenchel1980suspension, finlay1996spectacular} have been the focus of much study in both symbiotic and free-living contexts, and therefore provide an excellent model system for the study of alga-protist endosymbioses. 

The primary benefit of an endosymbiotic relationship between a heterotophic host and a photosynthetic symbiont (photosymbiosis) is thought to be nutrition \cite{johnson2011acquired, johnson2011acquisition, karakashian1963growth}. The host obtains nutrients from its environment via phagotrophy---the engulfing of cells or particles and subsequent digestion within a vacuole. Free-living algae are also ingested in this manner, but not all digested; rather, some resist digestion long enough for a section of the digestive vacuole membrane to `pinch off' and form a new, distinct vacuole. Known as the perialgal vacuole, this provides an alga with protection from digestion \cite{karakashian1973intracellular, kodama2011four}, allowing it to take up residence within the host ciliate and carry out the usual unicellular life cycle of growth and cytokinesis (division). Such symbiotic algae are now dependent on their host for nutrients which are unobtainable via photosynthesis (in particular, nitrogen). In return for these nutrients, the symbiont releases a portion of its photosynthate into the host cytoplasm, resulting in a net gain of organic carbon for the host \cite{brown1974transfer, muscatine1967soluble, ziesenisz1981evidence}.

The consequence of this nutrient exchange is that the photosymbiosis exists on a context-dependent continuum whereby the nature of the interaction depends on the light level \cite{lowe2015shining}. In low light, the correspondingly low photosynthetic output of the symbionts results in a net loss of nutrition for the host---this is effectively parasitism. As light increases, the increase in photosynthesis results in symbionts providing a net nutritional benefit to their host, yielding a mutualistic relationship.

A key step in the establishment of a permanent symbiotic relationship is the maintenance of a stable symbiont population \cite{muscatine1979regulation}. Clearly, if the host population grows more rapidly than the symbiont population, successive generations of host cells will become increasingly diluted until a completely aposymbiotic state is reached. This can occur when \emph{P. bursaria} are grown in the dark \cite{karakashian1963growth}. Conversely, if the symbiont population is the faster growing of the two, it will increase to the point of saturation, with severe consequences for the host---the symbionts have become parasites. Hence, if a host is to maintain a stable symbiont population it must carefully balance the gain and loss of symbionts so the two populations increase at approximately equal rates, either through controlling the rate of intake of new symbionts and the rate of removal (through expulsion or digestion), or by synchronising the cellular division cycles of the organisms. Both of these mechanisms could potentially lead to conflict between host and symbiont and the need for greater control by the dominant partner---presumed to be the host---to maintain the symbiosis.

Alga-protist endosymbioses have been addressed only briefly in the mathematical literature; see \cite{momeni2011using} for a review. There has been much emphasis on potential mechanisms for cell-cycle synchronisation in \emph{Chlorella}-\emph{Hydra} symbioses \cite{taylor1989maintenance, mcauley1990regulation} and \emph{Chlorella}-ciliate symbioses \cite{stabell2002ecological}, while others focus on the role of nutrient trading \cite{flynn2009building, hallock1981algal} in more general photosymbioses. Conditions determining the evolution of an obligate endosymbiosis were investigated in \cite{law1998symbiosis, weisbuch1993emergence, yamamura1996evolution}. A relevant recent paper modelled the \emph{P. bursaria}-\emph{Chlorella} symbiosis, showing how the combination of dynamic nutrient trading and differences in growth rates between partners yield a steady symbiotic population, but neglected to incorporate the potentially significant effects of symbiont intake and removal \cite{iwai2015maintenance}. To the best of our knowledge, the present work represents the first attempt to provide a comprehensive ecological model encapsulating the myriad facets of symbiosis across a range of environmental conditions.

In this article we develop a model to illustrate the mechanistic basis for a photosymbiotic relationship and the configuration of the resultant mixotrophic holobiont. We describe host-symbiont interactions by a deterministic system of ordinary differential equations, incorporating the vertical transmission of symbionts via host cytokinesis and the horizontal transmission of symbionts via ingestion from, and egestion into, the environment. The above discussion on host-symbiont cell-cycle synchronisation forms the basis of a key assumption; namely that on the timescale of our model, host and symbiont cytokinesis is almost concurrent, in that daughter cells have a symbiont load equal to that of their mother cell. This has been directly observed in \emph{P. bursaria} \cite{takahashi2007arrest}, and is in contrast to asynchronous cell cycles, for example, in which daughter cells have a symbiont load half that of their mother cell. The interplay between horizontal and vertical transmission of symbionts selects a particular symbiont distribution across the host population. We investigate how this distribution responds to different environmental conditions, in particular light levels, and how host control mechanisms may evolve. Note that our model is constructed in reference to the specific relationship between \emph{P. bursaria} and \emph{Chlorella}, but is readily reparameterised so as to be applicable to other photosymbioses. Hence we formulate our model using the language of a general symbiotic relationship between a heterotrophic host and a phototrophic symbiont, and parameterise it using data available in the literature on the \emph{P. bursaria}-\emph{Chlorella} symbiosis.

The paper is organised as follows. In Section \ref{Sec_Model} we derive our model. Incorporating nutrient trading and limitation allows us to formulate the host growth rate in terms of symbiont load and nutrient availability, highlighting the different strategies available to the host. This leads to the inclusion of general host control mechanisms, with particular attention paid to their impact on symbiont distribution via horizontal transmission. We then parameterise our model with respect to the specific \emph{P. bursaria}-\emph{Chlorella} relationship. In Section \ref{Sec_Equilibria} we perform numerical simulations of our model, highlighting the roles played by host control and irradiance in determining population equilibria. We then employ adaptive dynamics to illustrate how host control is able to evolve in Section \ref{Sec_AD}. We conclude by discussing our findings and intentions for future investigation in Section \ref{Sec_Discussion}. 


\section{The model}
\label{Sec_Model}
We describe the distribution of symbionts among the host population by defining the time-dependent set of variables $\bs{\phi} = (\phi_0, \phi_1, \ldots)^T$, where each entry $\phi_k(t)$, $k \in \mathbb{N}_0$, of the column vector $\bs{\phi}$ represent the concentration of hosts with $k$ symbionts at time $t$. We assume that the composition of the population changes due to the following processes:
\begin{itemize}
\item Cytokinesis of host cells, at rate $c_k$. We assume that the host and symbiont cell cycles are synchronised so as to be concurrent on the appropriate timescale; thus a host containing $k$ symbionts divides into two hosts that each contain $k$ symbionts. Also, we suppose that host cytokinesis is mediated by host population density.
\item Death of host cells, at rate $d_k$. Host death is independent of population density. 
\item Symbiont gain via ingestion of free-living potential symbionts, at rate $g_k$.
\item Symbiont loss, at rate $l_kk$, where we assume hosts lose symbionts at a rate proportional to their symbiont load.
\end{itemize}
Note that the first process encodes vertical transmission of symbionts, while the third and fourth encode horizontal transmission via the free-living population. 

Empirical evidence supports the hypothesis of cell cycle synchrony; Takahashi \emph{et al}. \cite{takahashi2007arrest} found that symbionts divided only when host cytoplasmic streaming ceased, which occurred just prior to host division. Each symbiont divided approximately once, resulting in two daughter cells with symbiont loads approximately equal to that of the mother before cessation of cytoplasmic streaming. This is opposed to, for example, a timescale-separated situation in which symbiont and host division is not concurrent, or a complete lack of synchrony as expected in an evolutionarily young symbiosis. In addition, we suppose that maintaining symbionts diverts nutrients away from cell growth, so that an excess of symbionts results in a net detrimental effect on the host (parasitism).

We have simplified the gain and loss processes to include only those ingestion events which result in the retention of a new symbiont and assume that loss of symbiont results in ejection from the cell. Although digestion of unwanted symbionts is more likely, for the sake of simplicity we do not explicitly account for this, nor for predation of free-living potential symbionts, in our model. We instead consider such effects to be sufficiently accounted for by the general feeding behaviour of the host (cf. Section 2.1).

Note that all rates depend on the symbiont number $k$, allowing us to explicitly incorporate the costs and benefits different symbiont loads bring to the host. Moreover, each of the two horizontal transmission processes yields a potential mechanism for host control of the symbiont population, by altering the gain and loss rates in response to the cost/benefit trade-off inherent to the symbiosis.

Taking all this into account, we see that the system varies according to the infinite set of ordinary differential equations
\begin{align}
\label{phi0_eqn}
\fr{\od \phi_0}{\od t} & = \lb[ c_0 \lb( 1 - \fr{\Phi}{K} \rb) - d_0 - g_0 \rb] \phi_0 + l_1 \phi_{1},
\\
\label{phik_eqn}
\fr{\od \phi_k}{\od t} & = g_{k-1} \phi_{k-1} + \lb[ c_k \lb( 1 - \fr{\Phi}{K} \rb) - d_k - g_k - l_kk \rb] \phi_k +l_{k+1} (k+1) \phi_{k+1},
\end{align}
where \eqref{phik_eqn} holds for all $k \in \mathbb{N}$ and we define the total host population
\begin{equation}
\Phi = \sum_{j=0}^\infty \phi_j.
\end{equation}
$K$ is the carrying capacity of the system. Note that if we define $\phi_{-1} \equiv 0$, setting $k=0$ in \eqref{phik_eqn} yields \eqref{phi0_eqn}. We write the equation for $\phi_0$ explicitly in order to emphasise the differing behaviour at the boundary $k=0$; in particular, that the only way for a host to leave the aposymbiotic state $\phi_0$ is to gain a symbiont via ingestion of an organism from the free-living population. 

\eqref{phi0_eqn}-\eqref{phik_eqn} can be written using matrix notation as
\begin{equation}
\label{phi_matrixeqn}
\fr{\od \bs{\phi}}{\od t} = \lb( A - \Phi B \rb)\bs{\phi},
\end{equation}
where the entries of the matrix $A$ are given by
\begin{equation}
\label{DefineA}
A_{k,j} = \lb\{ \begin{array}{lcl}
c_k - d_k - g_k - l_kk, & \quad & j = k,
\\
l_{k+1}(k+1), & \quad & j = k+1,
\\
g_k & \quad & j = k-1,
\\
0, & \quad & \mathrm{otherwise},
\end{array} \rb.
\end{equation}
and those of $B$ by
\begin{equation}
\label{DefineB}
B_{k,j} = \lb\{ \begin{array}{lcl}
c_k/K, & \quad & j = k,
\\
0, & \quad & \mathrm{otherwise},
\end{array} \rb.
\end{equation}
for $(k,j) \in \mathbb{N}_0^2$. Note we have separated \eqref{phi_matrixeqn} into linear and nonlinear parts, with the coefficients of the linear part yielding a tridiagonal matrix, $A$, and those of the nonlinear part yielding a diagonal matrix, $B$. 

The symbiotic relationship is characterised by a transfer of nutrients from the host to its symbionts in exchange for photosynthetically fixed carbon. Identifying carbon and nitrogen as the primary elements limiting growth, we describe this nutrient trading in the following manner. We assume that the host uses carbon and nitrogen in the ratio $\lambda_h$, and hence host growth rate is limited by
\begin{equation}
\min(C_k,\lambda_h N_k),
\end{equation}
where $C_k$ ($N_k)$ is the net intake rate of carbon (nitrogen) used directly by the host for growth, and is dependent upon symbiont number due to nutrient trading. We assume that the uptake of nutrients by phagotrophy (by the host) or photosynthesis (by the symbiont) occurs according to a Holling type II functional response \cite{fenchel1980suspension, dorling1997effect}. Each symbiont then yields a proportion $z_s \in [0,1]$ of its photosynthate to its host; in return, the $k$ symbionts receive a share of the nutrients (in particular, nitrogen) obtained by the host via phagotrophy. We assume that the total nutrition released by the host to its symbionts also varies according to a Holling type II functional response, with maximum $z_h \in [0,1]$; hence, as the symbiont load increases, the host gives up an increasing proportion of its phagotrophically obtained nutrition, but the share received by each individual symbiont decreases. We therefore have
\begin{align}
\label{Chk}
C_k & = C_F\lb( 1 - \fr{z_h k}{k+\kappa_h} \rb) + C_Lz_sk,
\\
\label{Nhk}
N_k & =  N_F\lb( 1 - \fr{z_h k}{k+\kappa_h} \rb),
\end{align}
where $C_F$ and $N_F$ are the amount of usable carbon and nitrogen provided by the hosts food, given by
\begin{align}
C_F & = \fr{\eta_C \lambda_F}{1+\lambda_F}\fr{bF}{F+\kappa_F},
\\
N_F & = \fr{\eta_N}{1+\lambda_F} \fr{bF}{F+\kappa_F},
\end{align}
and $C_L$ is the amount of usable carbon provided by symbiont photosynthesis, given by
\begin{equation}
\label{CL}
C_L = \eta_L \fr{aL}{L+\kappa_L}.
\end{equation}
Here $a$ represents the maximum symbiont photosynthesis rate and $b$ the maximum host phagotrophy rate; $F$ is the amount of bacterial food available to the host and $L$ the light level, each assumed constant; $\kappa_F$, $\kappa_L$ and $\kappa_h$ are the appropriate half-saturation constants; $\eta_L$, $\eta_C$ and $\eta_N$ are the conversion efficiencies of photosynthetically obtained carbon, phagotrophically obtained carbon and phagotrophically obtained nitrogen to cell growth; and $\lambda_F$ is the C:N ratio of the hosts food source. Thus the nutrient trading is characterized by the host and symbiont trading traits $z_h$ and $z_s$ and the half-saturation constant $\kappa_h$, relating to how a host distributes nutrients among its symbionts.

A host is carbon-limited if $C_k < \lambda_h N_k$, and nitrogen-limited if $C_k > \lambda_h N_k$. The conversion of nutrients into growth is therefore most efficient when $C_k - \lambda_hN_k = 0$ (cf. (\ref{Chk})-(\ref{Nhk})). We rearrange this condition to obtain
\begin{equation}
\label{klam_eqn}
z_sk^2 + \lb[ \kappa_h z_s + \mu\lb(1-z_h\rb) \rb] k + \mu\kappa_h = 0,
\end{equation}
where
\begin{equation}
\mu = \fr{C_F-\lambda_hN_F}{C_L}
\end{equation}
compares the $\lambda_h$-weighted nutrient intake via phagotrophy to the per capita symbiont carbon production (note that photosynthesis yields no nitrogen, and so the denominator contains only a carbon term). From the point of view of the host, $\mu<0$ represents carbon-deficient food and $\mu >0$ represents carbon-rich food. Roughly speaking, the closer $|\mu|$ is to zero, the more efficient phagotrophy is, in the sense that nutrient intake is closer to the optimal ratio $\lambda_h$. 

Equation (\ref{klam_eqn}) has at most one non-zero solution $k=k_\lambda$, where
\begin{equation}
\label{k_lambda}
k_\lambda = \fr{1}{2z_s} \lb( -\mu \lb(1-z_h\rb) - \kappa_hz_s + \sqrt{ \lb( \mu \lb(1-z_h\rb) + \kappa_hz_s \rb)^2 - 4\mu\kappa_hz_s} \rb).
\end{equation}
We can discount the negative square root as that solution of (\ref{klam_eqn}) is always negative (unless $\kappa_h=0$, in which case the negative square root yields $k_\lambda=0$; we assume $\kappa_h >0$ throughout). As $z_h, z_s \in [0,1]$, inspection of (\ref{k_lambda}) indicates that $k_\lambda$ is real and positive only if $\mu < 0$, i.e. if food is carbon-deficient. If $\mu=0$ then $k_\lambda=0$, representing the fact that symbionts provide no nutritional benefit when the host feeds on prey with a C:N ratio which is precisely that required for host growth. If $\mu > 0$ or $z_s = 0$ then (\ref{klam_eqn}) has no non-negative solutions and (\ref{k_lambda}) is physically meaningless; rather, the best strategy for a host in these cases is to divest itself of symbionts.

From inspection of $C_k$ (\ref{Chk}) and $N_k$ (\ref{Nhk}), we see that hosts are carbon-limited for $0\leq k < k_\lambda$, and nitrogen-limited for $k > k_\lambda$. Hence, if $k_\lambda < 0$ hosts are nitrogen-limited no matter their symbiont load. Note that $k_\lambda$ is in general not integer-valued, and so a perfect balance of nutrients according to the ratio $\lambda_h$ is in general unobtainable since the symbiont load $k$ is an integer. The best a host can do in practice is choose the symbiont load which minimises $C_k - \lambda_hN_k$, i.e. that for which $C_k - \lambda_hN_k$ is closest to zero, thus minimising any nutrient surplus. 

Associated with each symbiont there is also an upkeep cost, separate to nutrition, incorporating such processes as the maintenance of the perialgal vacuole, production of protein transporters, etc. For simplicity, we assume this cost to be proportional to symbiont load. Note that we discount the possibility that symbionts produce any benefits in addition to nutrition, in order to focus our attention on the nutrient trading. In light of the above discussion of nutrient trading and limitation, we therefore write the host cytokinesis rate as
\begin{equation}
\label{DefineHostCytokinesis}
c_k = \alpha_c \Rp\big( \min( C_k,\lambda_h N_k ) - Qk \big),
\end{equation}
where we define the ramp function
\begin{equation}
\Rp(y) := \lb\{ \begin{array}{lcl}
y, & \quad & y>0,
\\
0, & \quad & y \leq0. 
\end{array} \rb.
\end{equation}
$\alpha_c$ is the rate at which a host converts nutrition into growth (and therefore cytokinesis), while $Q$ is the additional upkeep cost per symbiont. Note that above a certain threshold symbionts become parasites, as the cost of upkeep becomes so burdensome to the host as to outweigh any nutritional benefits. Hence the host death rate is
\begin{equation}
\label{DefineHostDeath}
d_k = \alpha_c \Rp\big( Qk-\min( C_k,\lambda_h N_k ) \big) + \alpha_d,
\end{equation}
comprising the effects of an excessively burdensome symbiont load and the base death rate in the absence of symbionts, denoted by the constant $\alpha_d$. 

We can now define the net host growth rate as
\begin{equation}
\label{DefineHostGrowth}
r_k = c_k - d_k = \alpha_c \big( \min( C_k,\lambda_h N_k ) - Qk \big) - \alpha_d.
\end{equation}
Thus we see that a hosts symbiont load directly effects its potential for growth. In light of the analysis leading to the derivation of the optimum symbiont load $k_\lambda$ \eqref{k_lambda}, we have
\begin{equation}
\label{DifferenceInGrowthRate}
r_{k+1} - r_k = \lb\{
\begin{array}{lcl}
\alpha_c\lb( - C_F \dfrac{z_h \kappa_h}{(k+1+\kappa_h)(k+\kappa_h)} + C_Lz_s - Q \rb), & \quad & 0 \leq k \leq \lfloor k_\lambda \rfloor - 1,
\\
\\
\alpha_c \lb( \lambda_h N_{\lfloor k_\lambda \rfloor+1} - C_{\lfloor k_\lambda \rfloor} - Q \rb), & \quad & k = \lfloor k_\lambda \rfloor, 
\\
\\
-\alpha_c \lb( \lambda_h N_F\dfrac{z_h \kappa_h}{(k+1+\kappa_h)(k+\kappa_h)} + Q \rb), & \quad & k \geq \max(\lfloor k_\lambda \rfloor+1,0),
\end{array}\rb.
\end{equation}
where $\lfloor \cdot \rfloor$ denotes the integer part of a given real number and $k_\lambda$ is defined in \eqref{k_lambda}. When $k_\lambda < 0$, hosts are always nitrogen-limited ($C_k > \lambda_h N_k$ for all $k \in \mathbb{N}_0$) and $r_k$ is a monotonically decreasing function from $r_0$. In such a situation symbionts provide no benefit to the host. As our aim in the present article is to model photosymbiosis, we discount this situation as unrepresentative of the biological reality we are interested in, and shall not discuss it further.

Assuming $k_\lambda \geq 0$, we can identify three parameter ranges, in each of which the behaviour of $r_k$ as a function of $k$ is qualitatively distinct. From (\ref{DifferenceInGrowthRate}), we see that $r_k$ has a local maximum at $k = k_{\mathrm{max}}$, where
\begin{equation}
\label{kmax}
k_{\mathrm{max}} = \lb\{ \begin{array}{lcl}
\lfloor k_\lambda \rfloor, & \quad & Q > \lambda_h N_{\lfloor k_\lambda \rfloor+1} - C_{\lfloor k_\lambda \rfloor},
\\
\lfloor k_\lambda \rfloor \ \mathrm{and} \ \lfloor k_\lambda \rfloor+1, & \quad & Q = \lambda_h N_{\lfloor k_\lambda \rfloor+1} - C_{\lfloor k_\lambda \rfloor},
\\
\lfloor k_\lambda \rfloor + 1, & \quad & Q < \lambda_h N_{\lfloor k_\lambda \rfloor+1} - C_{\lfloor k_\lambda \rfloor},
\end{array} \rb.
\end{equation} 
and $C_k$ and $N_k$ are defined in \eqref{Chk}-\eqref{Nhk}. Furthermore, we can see that the first line of the right-hand side of (\ref{DifferenceInGrowthRate}) vanishes at $k=k_0$, defined as
\begin{equation}
\label{k0}
k_0 = - \kappa_h - \fr{1}{2} + \sqrt{\fr{1}{4} + \fr{C_Fz_h \kappa_h}{C_Lz_s - Q}};
\end{equation}
where $k_0$ is not necessarily integer-valued. Thus, provided $k_0$ is real and positive, $r_k$ has a local minimum at $k=k_{\mathrm{min}}$, defined as
\begin{equation}
\label{kmin}
k_{\mathrm{min}} = \lb\{ \begin{array}{lcl} 
\lb\lfloor k_0 \rb\rfloor + 1, & \quad & k_0 \notin \mathbb{N}_0,
\\
k_0 \ \mathrm{and} \ k_0+1, & \quad & k_0 \in \mathbb{N}_0.
\end{array}\rb.
\end{equation} 
Inspection of \eqref{DifferenceInGrowthRate} indicates that if $C_L z_s < Q$ then $r_k$ is monotonically decreasing; moreover, if $C_L z_s \geq Q$ then $k_{\mathrm{min}}$ is always real-valued. We shall henceforth assume the latter, as otherwise the per capita cost of maintaining the symbiosis outweighs the gain, in which case the model does not represent a biologically relevant relationship. 

\begin{figure}[t!]
\begin{center}
\includegraphics[width = \textwidth]{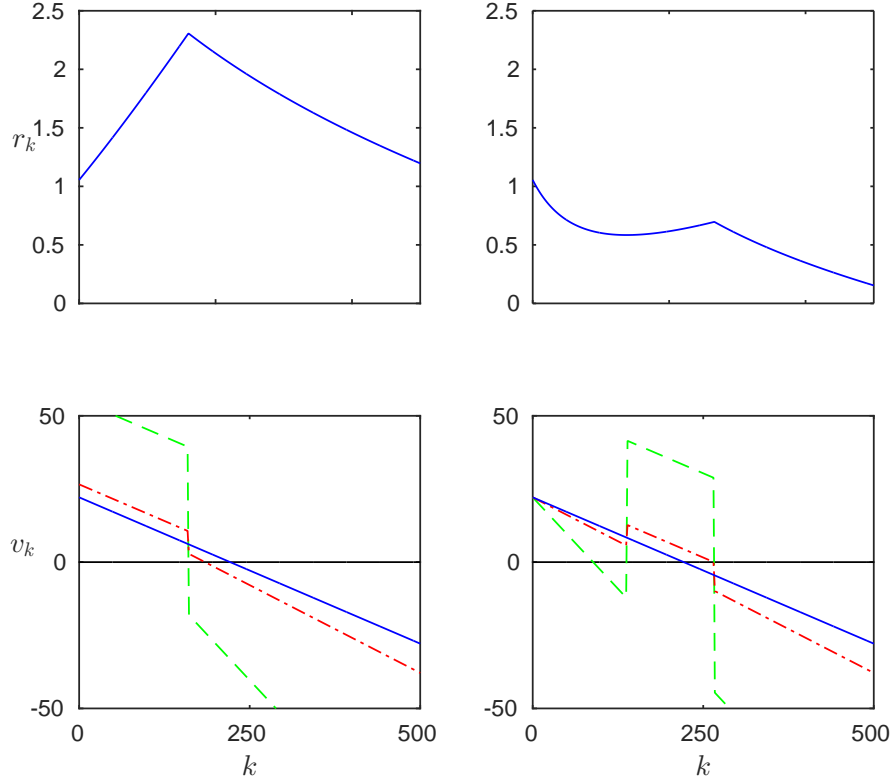}
\end{center}
\caption{\emph{Top: net host growth rate $r_k$ \eqref{DefineHostGrowth} for two qualitatively different host trading scenarios, $z_h=z_s=0.8$, $\kappa_h = 500$ (left) and $z_h=0.8$, $z_s=0.25$, $\kappa_h=100$ (right). Bottom: the associated approximate rate of horizontal transmission $v_k$ \eqref{Define_vk}, for three different values of host control, $\gamma=0$ (solid blue lines), $\gamma=0.2$ (dot-dashed red lines), $\gamma=1.5$ (dashed green lines). Other parameter values are given in table \ref{Table_ParameterValues}, with $L=20$.}}
\label{Fig_rv}
\end{figure}

We can therefore define the three regions in which $r_k$ exhibits qualitatively different behaviour as follows:
\begin{itemize}
\item $k_{\mathrm{min}} \geq k_{\mathrm{max}}$. Here $r_{k+1} > r_k$ for all $k \in \mathbb{N}_0$, yielding a host growth rate which is a monotonically decreasing function of $k$. Hence symbiont-free hosts exhibit the highest growth rate.
\item $0 < k_{\mathrm{min}} < k_{\mathrm{max}}$. In this region $k=k_{\mathrm{min}}$ defines a local minimum of $r_k$, with local maxima at $k=0, k_{\mathrm{max}}$. The existence of two maxima in the growth indicates a potential choice of strategy for the hosts, the precise details of which will depend heavily on horizontal transmission of symbionts.
\item $k_{\mathrm{min}} \leq 0$. $r_k$ now has a global maximum at $k=k_{\mathrm{max}}$, with a local minimum at $k=0$ and a global minimum at $k=\infty$. The optimal host strategy is a symbiont load of $k_{\mathrm{max}} \approx k_\lambda$.
\end{itemize}
The functional form of this growth rate for the biologically important region $k_{\mathrm{min}} \leq 0$, and the double-peaked scenario $0 < k_{\mathrm{min}} < k_{\mathrm{max}}$, can both be seen in figure \ref{Fig_rv}.

We now introduce the possibility of a host actively managing its symbionts over and above metabolic provision, as opposed to them being simply a passive load. In effect, the control of its symbiont load becomes a trait of the host. We suppose that control is implemented via the symbiont gain/loss terms in \eqref{phi_matrixeqn}, thus enabling hosts to bring about dynamic changes in the symbiont population, and that hosts choose to alter their symbiont load on the basis of increasing fitness, i.e. increasing $r_k$. This mathematical formulation serves as a proxy for the underlying physiological factors influencing the choice between investing in symbionts, or shedding them. 

We therefore define
\begin{equation}
\label{DefineGain}
g_k = \beta_g \fr{L}{L+\kappa_L}\lb\{ 1 + \gamma \Hv(r_{k+1}-r_k)\lb[ 1 - \fr{1}{2} \Hv(r_{k-1}-r_k) \rb]  \rb\},
\end{equation}
and
\begin{equation}
\label{DefineLoss}
l_k = \beta_l \lb\{ 1 + \gamma \Hv(r_{k-1}-r_k)\lb[ 1 - \fr{1}{2} \Hv(r_{k+1}-r_k) \rb] \rb\},
\end{equation}
where $\Hv(x)$ is the Heaviside step function with $\Hv(0)=0$ and we assign $r_{-1} = 0$. If $\gamma = 0$, there is no active host control and horizontal transmission of symbionts is a purely passive process. If $\gamma$ is non-zero, however, the host manages its symbiont load in order to increase its growth rate, increasing either its gain or loss rate as required in order to achieve this. The terms inside square brackets in (\ref{DefineGain})-(\ref{DefineLoss}) ensure that hosts do not get trapped at a local minimum of $r_k$; rather, hosts escape the minimum by either gaining or losing a symbiont, with equal proportions choosing each of the two strategies. In contrast, if a host finds itself at a local maximum it reverts to passive symbiont gain and loss only. If its symbiont load should change due to passive processes, host control kicks in once more to restore the optimal state. For simplicity, we assume hosts invest equally in both their gain and their loss rates, increasing each by an equal proportion of the respective passive values. We include the light-dependent factor in $g_k$ as a proxy for the free-living population of potential symbionts (cf. the rate of algal photosynthesis per cell in \eqref{CL}), which we assume constant with respect to time. This ensures that there is no ingestion of new symbionts in the dark, as purely phototrophic organisms cannot survive in the absence of light to drive photosynthesis. Our chosen formulation of the host control via a step increase in an appropriate parameter is probably the simplest from a mathematical point of view; however, it incorporates sufficient biological detail as to yield a useful means by which to investigate the phenomenon of host control.

It is clear from the structure of the discrete Burger's equation \eqref{phi_matrixeqn} that horizontal transmission is key in determining the symbiont distribution. This observation motivates the definition of the approximate rate of horizontal transmission
\begin{equation}
\label{Define_vk}
v_k = g_k-l_kk.
\end{equation}
Although this is an approximate formula only, it is nonetheless informative as to the qualitative behaviour of the symbiont distribution. Roughly speaking, if $v_k$ is positive then symbionts flow to the left, while if $v_k$ is negative they flow to the right. Of course, this picture is complicated by interplay between horizontal transmission and host population growth. However, one feature of \eqref{Define_vk} is especially useful; when $v_k$ decreases through zero, horizontal transmission acts to create a net flux towards this point from both directions, in effect creating an attractor for the symbiont dynamics. Thus we expect a peak to form at or near this point, depending on how strongly horizontal transmission is acting.

We plot $r_k$ \eqref{DefineHostGrowth} and $v_k$ \eqref{Define_vk} for two qualitatively different parameter regimes in figure \ref{Fig_rv}. In the scenario depicted by the left-hand panels, in which nutrient trading is relatively cheap for the host, $r_k$ has a single peak. Thus increasing the host control $\gamma$ from zero acts to shift $k_+$ closer to $k_{\mathrm{max}}$, the point at which $r_k$ is maximal, and at the same time increases the magnitude of the rate of horizontal transmission. On the other hand, the right-hand panels depict nutrient trading which is relatively expensive for the host, yielding a growth rate with two maxima. Increasing $\gamma$ from zero in this instance initially has the same qualitative effect as before. However, once $\gamma$ increases over a certain threshold then a second point appears at which $v_k$ decreases through zero. Thus we now have two possible attractors, separated by a repellor at which $v_k$ increases through zero. We discuss the consequences of this further in Section \ref{Sec_Equilibria}.

We denote the possible attractors, i.e. the points at which $v_k$ decreases through zero, by $k_+$. These can be found analytically. To this end, we define
\begin{equation}
\begin{split}
k_{10} &= \fr{L}{L+\kappa_L}\dfrac{\beta_g}{\beta_l}(1+\gamma),  
\\
k_{01} &= \fr{L}{L+\kappa_L}\dfrac{\beta_g}{\beta_l}\dfrac{1}{1+\gamma},
\\
k_{00} &= \fr{L}{L+\kappa_L}\dfrac{\beta_g}{\beta_l} = k_{11},
\end{split}
\end{equation}
to be the four possible (and not necessarily integer-valued) solutions of $v_k = 0$. The subscripts indicate which of the symbiont gain (first subscript) and loss (second subscript) processes are merely passive (subscript 0) or under host control (subscript 1) in the region of $k$-space that particular solution falls. Then $k_+$ can be determined by calculating whether $v_k$ is positive to the left and negative to the right of each of these solutions. Note we assume $\beta_g$ and $\beta_l$ are strictly positive throughout.

We note first that if $\gamma = 0$ then $k_+ = k_{00}$. For $\gamma > 0$, the situation is more complicated due to our formulation of the gain and loss rates \eqref{DefineGain}-\eqref{DefineLoss} in terms of the discontinuous Heaviside function. Assuming a single-peaked growth rate, so that $k_0 < 0$ (cf. \eqref{k0}), we have
\begin{equation}
\label{k+_SinglePeak}
k_+ = \lb\{ \begin{array}{lcl}
k_{10}, & \quad & 0 \leq k_{10} < k_{\mathrm{max}},
\\
k_{\mathrm{max}}, & \quad & k_{\mathrm{max}} = k_{00},
\\
k_{01}, & \quad & k_{01} > k_{\mathrm{max}};
\end{array} \rb.
\end{equation}
see the left-hand panels of figure \ref{Fig_rv}. If instead $k_0 \geq 0$, yielding a growth rate with two maxima, then
\begin{equation}
\label{k+_DoublePeak}
k_+ = \lb\{ \begin{array}{lcl}
k_{01}, & \quad & 0 < k_{01} < k_{\mathrm{min}},
\\
k_{10}, & \quad & k_{\mathrm{min}} < k_{01} < k_{\mathrm{max}},
\\
k_{\mathrm{max}}, & \quad & k_{\mathrm{max}} = k_{00},
\\
k_{01}, & \quad & k_{01} > k_{\mathrm{max}};
\end{array} \rb.
\end{equation}
see the right-hand panels of \ref{Fig_rv}. Note that $k_{\mathrm{min}}$ is never an attractor if $\gamma > 0$, and neither is zero if $\beta_g > 0$; hence we have omitted these possibilities from \eqref{k+_DoublePeak}. 

We have 
\begin{equation}
k_{01} \leq k_{00} \leq k_{10},
\end{equation}
for all physically realistic parameter values. We can therefore interpret the effects of increasing host control as follows. When $\gamma=0$, both \eqref{k+_SinglePeak} and \eqref{k+_DoublePeak} simplify to $k_+ = \max(k_{00},0)$, i.e. horizontal transfer is strictly passive. As $\gamma$ increases from zero, $k_{10}$ increases while $k_{01}$ decreases. Thus $k_+$ approaches the local maxima of $r_k$ as $\gamma$ increases, with $k_+$ double-valued if $0 < k_{\mathrm{min}} < k_{\mathrm{max}}$. We note there is a third possibility, in which $k=0$ is a global maximum; in this case increasing host control pushes $k_+$ closer to zero, in effect destroying the symbiosis. We dismiss this case as being biologically uninteresting, and mention it only for completeness.

We can therefore see that there are four possible outcomes of the population dynamics:
\begin{itemize}
\item Stable symbiosis. $k_+$ is single-valued; the peak drifts until it reaches $k_+$, where it remains. 
\item Dichotomous symbiosis. $k_+$ is double-valued. Hosts may potentially split between the two strategies of low and high symbiont loads.
\item Near loss of symbionts. This corresponds to the limit $k_+ \rightarrow 0$ as $\beta_g/\beta_l \rightarrow 0$. In practice, a large but finite $\beta_l$ results in a very low but non-vanishing symbiont load.
\item Extinction due to symbiont overload. This corresponds to the limit $k_+ \rightarrow 0$ as $\beta_g/\beta_l \rightarrow \infty$. Gain of symbionts is sufficiently higher than loss to push the population into the region of negative growth ($r_k <0$), at which point the cost of maintaining symbionts is too much and the hosts die, leading to host population crash.
\end{itemize}
We shall focus on the first two outcomes as biologically relevant, with particular attention paid to the stable, single-peaked symbiosis.

We have estimated the model parameters from data available in the literature. We summarise the values used in table \ref{Table_ParameterValues}, and outline the rationale behind each choice below. Some data have been converted into the units used in the present work; we omit the details where this is trivial, e.g. hours to days. All parameters are given to two significant figures.

\begin{itemize}
\item C:N ratios. Finlay and Uhlig (\cite{finlay1981calorific}, table 1) calculated the elemental composition of various protozoa, of which the closest relative to \emph{P. bursaria} is \emph{P. caudatum}, which has a C:N ratio of 3.5 by weight, or 4.1 by amount of substance. They also found the C:N ratio by weight of a freshwater bacterial sample to be 3.9, or 4.6 by amount of substance. This is in agreement with the value given by Fagerbakke et al (\cite{fagerbakke1996content}, table 2) for various native aquatic and cultured bacteria.

\item Conversion efficiencies. Herrig and Falkowski \cite{herrig1989nitrogen} found the conversion efficiency of photosynthetically produced carbon into growth by algae to be 85\%, which we shall use for $\eta_L$. Jones et al \cite{jones2002effect} found that nitrogen growth efficiency was around 2.5 times the carbon growth efficiency of a marine copepod when nitrogen was in short supply, and so intake is maximised at the expense of efficiency. As hosts must uptake enough nitrogen for themselves and their symbionts, we expect nitrogen intake to be similarly prioritised in our system. Furthermore, Mauclaire \emph{et al}. \cite{mauclaire2003assimilation} found that 12\% of predated bacterial carbon was transformed into protist biomass in a bacteria-protist predator-prey system. These data lead us to assign the values $\eta_C = 0.12$ and $\eta_N = 0.3$ to the phagotrophic conversion efficiencies.

\item Photosynthesis. In \cite{finlay1996spectacular}, Finlay \emph{et al}. analysed  symbiont photosynthesis in a naturally occurring population of mixotrophic ciliates. In particular, they fitted photosynthetic data to a Holling type II function as in (\ref{CL}), calculating the maximum rate of photosynthesis to be $123$ $\mu$mol C mg$^{-1}$ Chl $a$ h$^{-1}$ and the slope of the light-response curve at low light to be $17.4$ $\mu$mol C mg$^{-1}$ Chl $a$ h$^{-1}$ $(\mu$mol photon m$^{-2}$ s$^{-1})^{-1}$ (\cite{finlay1996spectacular}, figure 3). Using the value given in the text of $0.4$ pg chlorophyll $a$ per \emph{Chlorella} and converting to units of d$^{-1}$, we therefore arrive at $a = 1.2 \times 10^{-6}$ $\mu$mol C d$^{-1}$, and $\kappa_L = 7.1$ $\mu$mol photon m$^{-2}$ s$^{-1}$. We consider irradiances $L$ in the range 0--50 $\mu$mol photon m$^{-2}$ s$^{-1}$.

\item Phagotrophy. Fenchel \cite{fenchel1980suspension} performed a comprehensive investigation of suspension feeding in ciliates, including data from the two \emph{Paramecium} species \emph{P. caudatum} and \emph{P. trichium}, which we shall use in lieu of data on \emph{P. bursaria}. Figure 2 of \cite{fenchel1980suspension} provides an estimate for the maximum ingestion rate as $4.32\times 10^{5}$ $\mu$m$^3$ d$^{-1}$. Table 2 in \cite{fagerbakke1996content} yields the estimated values for the cellular C:volume ratio of $8.3 \times 10^{-9}$ $\mu$mol $\mu$m$^{-3}$. As the C:N ratio is $\lambda_F = 14/3$, the proportion of food by amount of substance which is carbon is $\lambda_F/(1+\lambda_F) = 14/17$. Hence $b = (17/14) \times 4.3 \times 8.3 \times 10^{-4} = 4.4 \times 10^{-3}$ $\mu$mol d$^{-1}$. Figure 6 in \cite{fenchel1980suspension} provides a value of $4\times 10^7$ $\mu$m$^3$ ml$^-1$ for the half saturation constant. We estimate the volume of a bacterial cell to be 2$\mu$m$^3$ from the data in table 1 of \cite{fagerbakke1996content}, yielding $\kappa_F = 2\times 10^7$ cell ml$^{-1}$. We take the bacterial food concentration to be $5.8 \times 10^6$ cell ml$^{-1}$.

\item Host growth. The precise values of $\alpha_c$, $\alpha_d$ and $Q$ are difficult to determine, as we have formulated our model in terms of a growth rate $r_k$ dependent on individual symbiont loads, whereas data is calculated at the population level. However, the data of Karakashian \cite{karakashian1963growth}, who carried out a series of growth experiments in differing environmental conditions, yields enough information for us to make reasonable estimates. Bleached \emph{P. bursaria} grown in sterile media were found to simply die due to the complete lack of available nutrition, providing an estimate for the base death rate $\alpha_d = 0.7$ d$^{-1}$ (\cite{karakashian1963growth}, figure 1). Green \emph{P. bursaria} grown with abundant food and light had a mean daily fission rate of around 1.9 d$^{-1}$ (\cite{karakashian1963growth}, table 2). Furthermore, we know from the analysis around \eqref{k0} that for symbiosis to be desirable the upkeep cost $Q$ per symbiont must be less than the nutritional benefit per symbiont. Combining these considerations leads us to choose $\alpha_c = 1.8\times 10^4$ cell $\mu$mol nutrient$^{-1}$ and $Q = 5 \times 10^{-8}$ $\mu$mol nutrient$^{-1}$ cell$^{-1}$ d$^{-1}$. We choose $K = 1000$ cell ml$^{-1}$ as representative of a typical carrying capacity.

\item Nutrient trading. $z_h$, $z_s$ and $\kappa_h$ are traits we vary in order to investigate the symbiosis, and so we shall not fix them at any particular value.

\item Horizontal transmission. In a reinfection experiment \cite{karakashian1973intracellular}, Karakashian reported that aposymbiotic \emph{P. bursaria} exposed to a highly concentrated algal population ingested a mean of 285 in 2.5 mins, and after 23.5 hours a mean of 60 of these remained. We shall therefore take 60d$^{-1}$ as the maximum ingestion rate, as we assume that if the hosts had means and motive to ingest more symbionts then more than 60 would have remained. We shall assume that, as experimental irradiance was high, host control was in effect, and so the passive ingestion rate $\beta_g$ is lower than 60 d$^{-1}$; we choose $\beta_g = 30$ d$^{-1}$ and vary $\gamma$ from zero. Data appears to be scarce to nonexistent for rates of symbiont loss. We therefore simply assume a low passive loss of $\beta_l = 0.1$d$^{-1}$.

\end{itemize}

\begin{table}[t!]
\begin{center}
\begin{tabular}{|c|c|c|c|c|}
\hline
Process & Parameter & Dimensions & Value(s) & Source
\\
\hline
C:N ratios & $\lambda_h$ & mol C mol N$^{-1}$ & 4.1 & \cite{finlay1981calorific}
\\
& $\lambda_F$ & mol C mol N$^{-1}$ & 4.6 & \cite{fagerbakke1996content}
\\
\hline
Conversion & $\eta_L$ & 1 & 0.85 & \cite{herrig1989nitrogen}
\\
efficiencies & $\eta_C$ & 1 & 0.12 & \cite{jones2002effect, mauclaire2003assimilation}
\\
& $\eta_N$ & 1 & 0.3 & \cite{jones2002effect, mauclaire2003assimilation}
\\
\hline
Photosynthesis & $a$ & $\mu$mol C cell$^{-1}$ d$^{-1}$ & $1.2\times 10^{-6}$ & \cite{dorling1997effect}
\\
& $\kappa_L$ & $\mu$mol photon m$^{-2}$ s$^{-1}$ & 7.1 & \cite{dorling1997effect}
\\
& $L$ & $\mu$mol photon m$^{-2}$ s$^{-1}$ & 0--50 & n/a
\\
\hline
Phagotrophy & $b$ & $\mu$mol nutrient d$^{-1}$ & $4.4\times10^{-3}$ & \cite{fenchel1980suspension}
\\
& $\kappa_F$ & cell ml$^{-1}$ & $2\times10^7$ & \cite{fenchel1980suspension}
\\
& $F$ & cell ml$^{-1}$ & $5.8\times10^6$ & n/a
\\
\hline
Nutrient & $z_h$ & 1 & 0--1 & n/a
\\
trading & $z_s$ & 1 & 0--1 & n/a
\\
& $\kappa_h$ & cell & 0--$\infty$ & n/a
\\
\hline
Host growth & $\alpha_c$ & $\mu$mol nutrient$^{-1}$ & $1.8\times10^4$ & \cite{karakashian1963growth}
\\
& $\alpha_d$ & d$^{-1}$ & $0.7$ & \cite{karakashian1963growth}
\\
& $Q$ & $\mu$mol nutrient cell$^{-1}$ d$^{-1}$ & $5\times10^{-8}$ & \cite{karakashian1963growth}
\\
& $K$ & cell ml$^{-1}$ & 1000 & n/a
\\
\hline
Horizontal & $\beta_g$ & d$^{-1}$ & 30 & \cite{karakashian1973intracellular}
\\
transmission & $\beta_l$ & cell$^{-1}$ d$^{-1}$ & 0.1 & n/a
\\
& $\gamma$ & 1 & 0--$\infty$ & \cite{karakashian1973intracellular}
\\
\hline
\end{tabular}
\end{center}
\caption{\small{\emph{Model parameters and their numerical values, with reference to empirical data where possible.}}}
\label{Table_ParameterValues}
\end{table}



\section{Population equilibria}
\label{Sec_Equilibria}

We now present numerical solutions of the model \eqref{phi_matrixeqn}, highlighting the key ecological processes leading to population equilibria. We truncate the system at $k = k_{\mathrm{trunc}}$, where $k_{\mathrm{trunc}}$ is chosen so that $\phi_{k_{\mathrm{trunc}}}$ is exponentially small and therefore has a negligible effect on the solution near the symbiont peak. For the present purposes, setting $k_{\mathrm{trunc}} = 500$ turns out to be sufficient; although the growth rate $r_{500}$ is usually positive for the parameter values in table \ref{Table_ParameterValues}, horizontal transmission is sufficiently negative as to render population equilibria in which $\phi_{500} \approx 0$. We calculate steady-state solutions of the nonlinear problem \eqref{phi_matrixeqn} using Newton-Raphson iteration. We also calculate the dominant eigenvalues of the associated linear problem and the dominant eigenvectors of the associated linear problem $\dot{\bs{\phi}} = A\bs{\phi}$, i.e. the eigenvector of $A$ with eigenvalue of largest real part of all eigenvectors with no negative components.

\begin{figure}[t!]
\begin{center}
\includegraphics[width = \textwidth]{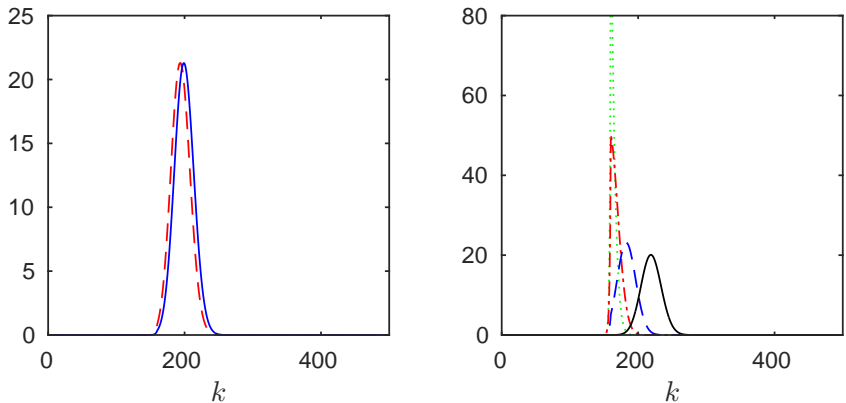}
\end{center}
\caption{\emph{Typical solutions for a single-peaked growth rate distribution $r_k$ (cf. figure \ref{Fig_rv}). Left: steady-state solution of \eqref{phi_matrixeqn} (blue solid line) and dominant eigenvector of the associated linear problem (red dashed line) with $\gamma = 0.1$. The eigenvector is scaled so its maxima is equal to the maxima of the steady state. Right: steady-state solutions of \eqref{phi_matrixeqn} for increasing host control, with $\gamma = 0$ (solid black line), $\gamma = 0.2$ (dashed blue line), $\gamma = 0.4$ (dot-dashed red line) and $\gamma = 0.6$ (dotted green line). Other parameter values are given in table \ref{Table_ParameterValues}, with $z_h=z_s=0.8$, $\kappa_h=500$ and $L=20$.}}
\label{Fig_Control}
\end{figure}

In figure \ref{Fig_Control} we present steady state solutions of \eqref{phi_matrixeqn} with a single-peaked growth rate distribution, corresponding to the left-hand panels of figure \ref{Fig_rv}. We see that the solution is closely approximated by the dominant eigenvalue of the associated linear problem. In fact, our numerical investigations showed that this agreement is excellent for a wide range of relevant parameter values, suggesting horizontal transmission rapidly organises the symbiont distribution. Furthermore, the location of the peak is approximately given by $k_+$ (not shown for the sake of clarity; cf. \eqref{k+_SinglePeak}), indicating that $v_k$ \eqref{Define_vk} is an excellent description of the rate of horizontal transmission. We can also see how increasing host control strength $\gamma$ moves the symbiont peak closer to the optimal value $k_\lambda$, and reduces its variance, indicating that hosts are maximising their benefit from the symbiotic relationship as $\gamma$ increases.

\begin{figure}[t!]
\begin{center}
\includegraphics[width = \textwidth]{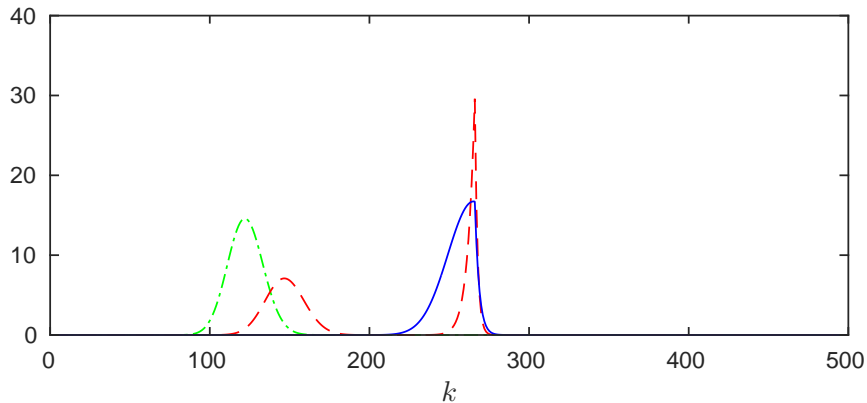}
\end{center}
\caption{\emph{Typical solutions for a double-peaked growth rate distribution $r_k$ (cf. figure \ref{Fig_rv}) and increasing host control, with $\gamma = 0.2$ (solid blue line), $\gamma = 0.5$ (dashed red line) and $\gamma = 0.8$ (dot-dashed green line). Other parameter values are given in table \ref{Table_ParameterValues}, with $Q = 1\times10^-7$, $z_h=0.8$, $z_s=0.25$, $\kappa_h=100$ and $L=20$.}}
\label{Fig_DoublePeak}
\end{figure}

In figure \ref{Fig_DoublePeak} we present steady-state solutions for a double-peaked growth rate distribution, corresponding to the right-hand panels of figure \ref{Fig_rv}, although we note that the per capita cost $Q$ in figure \ref{Fig_DoublePeak} is double that of figure \ref{Fig_rv}. Again, agreement between steady state solutions and dominant eigenvectors is excellent, although we omit the eigenvectors for clarity. For low $\gamma$, the peak is near $k_\lambda$; as $\gamma$ increases, the solution becomes double-peaked before losing the upper peak altogether as $\gamma$ increases further. In this case, increasing $\gamma$ without bound will have the effect of destroying the symbiosis, as increased host control moves the lower peak ever closer to zero. We note that not all double-peaked growth rates, or even double-valued $k_+$, yield corresponding double-peaked symbiont distributions. The deciding factor appears to be the net flux towards an attractor; the upper peak is usually at an advantage in this respect as hosts originating to the right of $k_{\mathrm{max}}$ are attracted to the upper peak and hence cannot reach the lower. 

\begin{figure}[t!]
\begin{center}
\includegraphics[width = \textwidth]{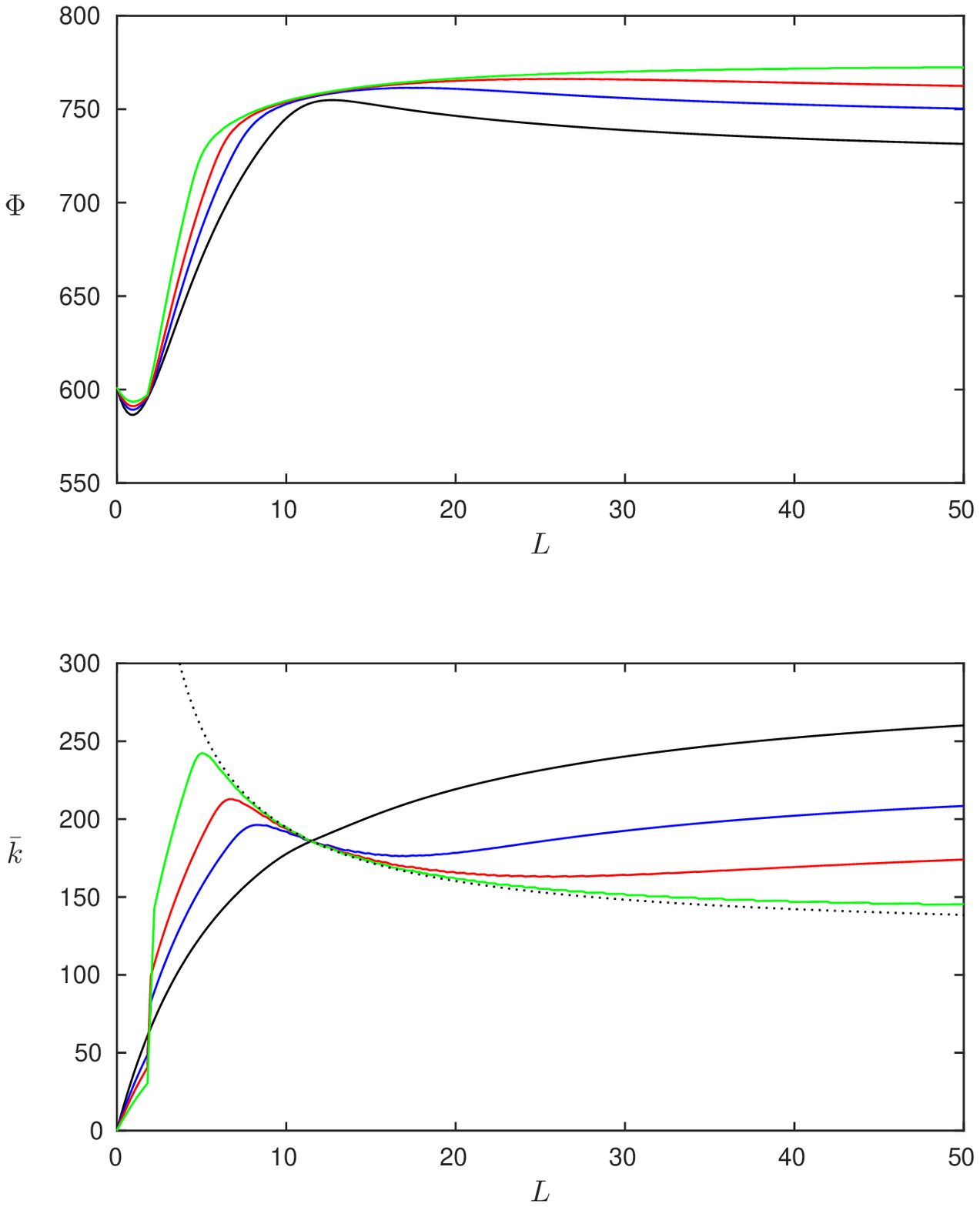}
\end{center}
\caption{\emph{Effects of irradiance and host control on host and symbiont populations. Above: total host population. Below: mean symbiont load. The dotted black line indicates the (non-integral) optimum symbiont load $k_\lambda$ \eqref{k_lambda}. Host control strengths are $\gamma = 0$ (black), $\gamma = 0.25$ (blue), $\gamma = 0.5$ (red) and $\gamma = 1$ (green). $z_h = z_s = 0.8$, $\kappa_h = 500$.}}
\label{Fig_LightResponse}
\end{figure}

A key environmental property affecting the symbiosis is light. In figure \ref{Fig_LightResponse} we plot the total host population $\Phi$ and the mean symbiont load
\begin{equation}
\bar{k} = \fr{1}{\Phi}\sum_{j=0}^\infty j\phi_j,
\end{equation}
of steady state solutions to \eqref{phi_matrixeqn}, for four levels of host control. We note that in each case the agreement between the numerically computed values of $\bar{k}$ and the approximate prediction $k_+$ is excellent, but neglect to plot this information to preserve clarity. We see in figure \ref{Fig_LightResponse}b that when there is no host control, the mean symbiont number simply increases with light according to the functional response incorporated into $g_k$ (cf. (\ref{DefineGain})). With a moderate level of host control, the mean symbiont number increases from zero to a local maximum, at which point it decreases in line with $k_\lambda$ before increasing again as passive gain overcomes host control due to an increasing free-living population. Strong enough host control prevents this, ensuring optimal nutrient intake at all light levels above a threshold, correlating with the observed increase in host population with higher levels of host control. We note that there is an initial dip in host population as irradiance increases from zero, indicating that at first symbionts are not photosynthesising enough to provide a net benefit to their hosts. Increased host control reduces this dip as the hosts invest in expelling unwanted symbionts. Note the sharp increase in $\bar{k}$ at low light; the solution changes qualitatively at this point, with a discontinuous change in the location of the peak.

\section{Adaptive dynamics}
\label{Sec_AD}

Having discussed the ecological aspects of the model in detail, we now turn our attention to the evolution of the symbiosis. In particular, we shall investigate the evolution of host control by allowing  the strength of control, $\gamma$, to evolve in an adaptive dynamics framework. The benefits of increased host control are likely to be balanced by an associated cost, which we model by taking the per capita cost of symbiosis maintenance to be a monotonically increasing function of $\gamma$, specifically 
\begin{equation}
Q \equiv Q(\gamma) = Q_0 + \fr{Q_1\gamma}{Q_2+\gamma},
\end{equation}
where the $Q_i$, $i = 1,2,3$, are constants. Thus the cost of symbiosis increases from an initial value of $Q_0$ at $\gamma=0$ to a maximum of $Q_0 + Q_1$ at $\gamma = \infty$, with $Q_2$ determining the rate at which this transition occurs. This limiting behaviour of the per symbiont cost provides a trade-off curvature as the trait evolves.

The adaptive dynamics approach assumes a separation of timescales between ecological and evolutionary dynamics, in that the ecological equilibrium is continually updated in evolutionary time by a series of successful invasions by initially rare mutants. A host with a mutant trait $\gamma'$ must be able to overcome the background state provided by the resident steady state population with trait $\gamma$. Hence the initial dynamics of a newly-appeared mutant $\bs{\psi}(t)$ are governed by 
\begin{equation}
\label{ADEqn}
\fr{\od \psi_k}{\od t} \approx g_{k-1}'\psi_{k-1} + \lb[ c_k'\lb( 1 - \fr{\Phi^*}{K} \rb) - d_k' - g_k' - l_k'k \rb] \psi_k + l_{k+1}'(k+1)\psi_{k+1},
\end{equation}
where $\Phi^* = \sum_{j=0}^\infty \phi_j^*$ and the $\phi_j^*$ are a steady-state solution of (\ref{phi_matrixeqn}) in the absence of mutants, i.e.
\begin{equation}
\label{SteadyState}
0 = g_{k-1}\phi_{k-1}^* + \lb[ c_k\lb( 1 - \fr{\Phi^*}{K} \rb) - d_k - g_k - l_kk \rb] \phi_k^* + l_{k+1}(k+1)\phi_{k+1}^*.
\end{equation}
Note that unprimed rate functions refer to the resident population with trait $\gamma$, and primed variables refer to the mutant population with trait $\gamma'$. 

We can see that (\ref{ADEqn}) is an eigenvalue problem. Therefore, if there exists an eigenvalue with positive real part corresponding to an eigenvector with no negative elements, the initially rare mutant population will exhibit exponential growth until nonlinear effects become significant. We shall make the usual adaptive dynamics assumption that such an invasion is always successful, with mutants replacing residents and achieving a new equilibrium, distinct from $\bs{\phi}^*$ due to the updated trait value $\gamma'$.

\begin{figure}[t!]
\begin{center}
\includegraphics[width = \textwidth,]{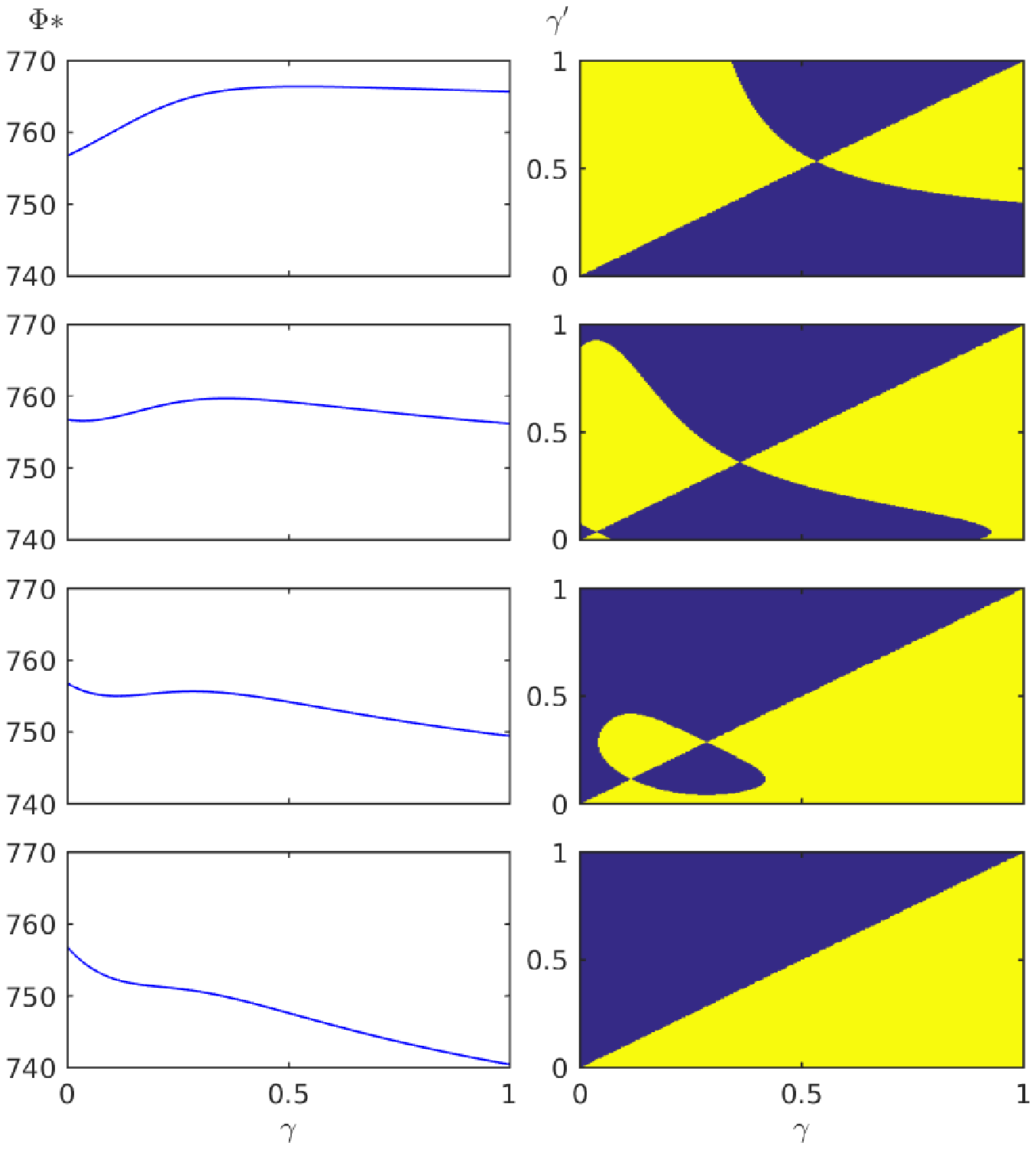}
\end{center}
\caption{\emph{Adaptive dynamics of \eqref{phi_matrixeqn}, for increasing values of the maximum cost $Q_0+Q_1$. Left-hand column: solution curves of \eqref{SteadyState}. Right-hand column: corresponding pairwise invasibility plots. Yellow regions indicate resident trait values $\gamma$ which are invasible by the corresponding mutant trait $\gamma'$, while blue regions are uninvasible. From the top, the panels on the first row were calculated with $Q_1 = 5\times10^{-8}$, the second with $Q_1 = 11\times10^{-8}$, the third with $Q_1 = 15\times10^{-8}$ and the fourth with $Q_1 = 20\times10^{-8}$. We take $Q_0 = 2\times 10^{-8}$ and $Q_2 = 0.5$ in all four calculations. Other parameter values are as in table \ref{Table_ParameterValues}, with $z_h=z_s=0.8$, $\kappa_h=500$ and $L=20$.}}
\label{Fig_AD}
\end{figure}

The eigenvalues of \eqref{ADEqn} must be calculated numerically. To this end, we first found the solution curve comprising solutions to \eqref{SteadyState} for different values of the resident trait $\gamma$ across the desired range. Then, for each resident trait we calculated the eigenvalues of \eqref{ADEqn}, for each value of the mutant trait $\gamma'$ across the same range. Following the adaptive dynamics methodology, we differentiate between the regions in which at least one eigenvalue corresponding to an eigenvector with no negative components has positive real part, and those in which none do. The former regions of trait space are therefore those in which a successful invasion may occur. We plot our results  in figure \ref{Fig_AD} for various values of $Q_1$, taking $Q_0 = 2\times 10^{-8}$ and $Q_2 = 0.5$. We see that if the cost increases relatively gradually with $\gamma$, a moderate level of host control will evolve. As $Q_1$ increases, an evolutionary steady state which is not convergence stable emerges from the origin, resulting in a barrier to the evolution of host control; if a mutation appears which is large enough, however, there still exists an evolutionary stable strategy with $\gamma \neq 0$. As $Q_1$ increase further, the two evolutionary steady states approach one another and coalesce before vanishing, in which case evolution favours no host control. We note that evolutionary stable strategies are associated with host population maxima, and evolutionary unstable strategies with population minima.

\section{Discussion}
\label{Sec_Discussion}

We have constructed a deterministic model of the endosymbiosis between a heterotrophic host and a phototrophic symbiont. By formulating the host growth rate in terms of its nutritional state, we were able to explicitly incorporate nutrient trading between a host and its symbionts into our model, allowing us to investigate the consequences of symbiosis for host fitness. A novel feature of our model is an explicit mechanism by which the host can exert a degree of control over the number of endosymbionts via adjustment of the rates of loss and gain of symbionts. The strength of this host control is captured by a single parameter in our model.

Host control is an essential addition to a model of this symbiosis in order to describe the behaviour of holobiont in response to variable levels of light. The model presented here predicts that the optimal symbiont load $k_\lambda$ \eqref{k_lambda} decreases monotonically with irradiance from infinity at zero light. This is because fewer symbionts are needed for the same gain as photosynthetic output increases, but this gain can only be realised if the host divests itself of excess symbionts, thus diverting nitrogen from symbiont nutrition back to its own growth. Such a response requires a level of host control over its symbiont load---without it, the symbiont population will increase with light to the detriment of the host. Thus, as light increases from zero, the symbiotic relationship moves from parasitism through commensalism to exploitation of one party by the other, depending on the level of control exerted by the host. 

The model leads to the following predictions at different light levels. When environmental irradiance increases from zero, after an initial dip in which symbiosis is slightly detrimental to the host, we observe a sharp increase in symbiont load. This is mainly due to an increase in ingestion as the free-living population of algae increases from zero in the dark. As light levels continue to increase, at a certain irradiance the mean symbiont load coincides with the optimal load $k_\lambda$. According to the level of explicit control imposed by the host on the symbiont, the model predicts that either the symbiont load will continue to increase, proportional to the rate of photosynthesis, or will remain at $k_\lambda$ and therefore decrease with increasing light. If host control is strong enough, this decrease is monotonic and the mean symbiont load remains approximately equal to $k_\lambda$. However, for weaker levels of host control, the symbiont load once again begins to increase due to the free-living population becoming too large for the host to effectively manage (cf. figure \ref{Fig_LightResponse}). Thus the differing strategies at higher light levels introduce conflict between host and symbionts, with the possibility that symbionts may counter-adapt defences against host control. Our model therefore demonstrates that adaptation to different light levels may not lead to easily predictable results, as coevolution may drive the symbiosis to differing regimes. In particular, strains adapted to higher light levels should exhibit a greater level of host control in order to overcome the increased symbiont gain due to increased growth of free-living algae and prevent exploitation of the hosts by their symbionts. 

This pattern of behaviour is precisely seen in recent experiments on this host-symbiont relationship. Lowe \emph{et al}. \cite{lowe2015shining} show that a peak in symbiont load is observed at a relatively low irradiance, and the symbiont load then decreases monotonically as the light levels increase. In our model this suggests that the symbiotic relationship of the strain studied has evolved to have a high level of host control. To confirm that such an evolutionary end point is a likely result of our model we have presented an adaptive dynamics approach using the host control strength $\gamma$ as the evolvable trait. The control is traded off against the per symbiont cost to the host, representing the investment of the host in explicit mechanisms to restrain the symbiont -- nutrient consumption is explicitly accounted for elsewhere in the model. Our evolutionary model demonstrates that when the cost to the host of imposing control is low then it is favourable to adopt a high level of host control; there is a convergent and evolutionary stable state for a finite value of $\gamma$. However, the fine detail of this evolutionary end point is sensitive to the per-capita costs of control, which suggests that different strains are likely to have evolved different levels of host control and this should be reflected in their response to a light gradient. 

If the per capita cost $Q$ increases too severely with the evolution of control then the convergent and evolutionary stable state is $\gamma = 0$, i.e. hosts do not evolve control. In this parameter regime our adaptive dynamics description no longer encapsulates the essential evolutionary pressures, with evolution beginning to act on the passive gain and loss rates via the parameters $\beta_g$ \eqref{DefineGain} and $\beta_l$ \eqref{DefineLoss}, for example. A likely end result of such evolution is the transition back into a regime in which conditions are favourable for the evolution of host control.

There exists also a more drastic scenario which we have not covered in Section \ref{Sec_AD}. If $Q$ increases sufficiently it can cause a transition from the solution branch in which the symbiont peak is near $k_\lambda$ to that with a peak to the left of $k_0$. In this case, increasing host control acts to destroy the symbiosis as the symbiont peak moves towards $k=0$ (cf. the discussion around figure \ref{Fig_DoublePeak} in Section \ref{Sec_Equilibria}). Furthermore, $\gamma$ is then free to increase without bound until the symbiosis is completely destroyed. However, once again we find our adaptive dynamics description failing to capture all relevant evolutionary effects; in this case, we expect ever increasing investment in host control to begin to have a detrimental effect on host growth in other ways than only increasing $Q$, for example decreasing the cytokinesis rate $\alpha_c$ \eqref{DefineHostCytokinesis} or increasing the death rate $\alpha_d$ \eqref{DefineHostDeath}.

Double-peaked growth rate distributions, in which the host has a choice of strategies, occur when nutrient trading is expensive for the host, thus reducing its benefit from the symbiosis. In reinfection experiments \cite{karakashian1973intracellular, kodama2012cell}, it is often observed that the resulting symbiont load distribution is double-peaked. Our model suggests that this could be due to a decreased release of photosynthate from new symbionts to their hosts, potentially a result of the time spent living autonomously, or perhaps due to a delay between ingestion of symbionts and commencement of nutrient trading. Moreover, our results indicate it is likely that one peak is transient, and given enough time, will decay leaving only a single peak. As discussed above, the double-peaked growth rate distribution provides a mechanism, as Q increases, whereby it is possible for the evolutionary system to select a non-symbiotic state, but without explicit modelling of the symbiont dynamics and ecology this remains a speculation.

The construction of  a more realistic mechanism by which cell-cycles become synchronised, by incorporating nutritional dependence into symbiotic growth rates, is a desirable feature currently lacking in our model. This also permits the inclusion of other host control mechanisms, implicit here, whereby the host manages the nutrient supply to its symbionts in order to maintain a stable population, perhaps including such dynamic nutrient trading as that described in \cite{iwai2015maintenance}. The precise nature of cell-cycle synchronisation is at present poorly understood, and including such effects in our more comprehensive model would build greatly on the simpler approaches attempted here and previously \cite{taylor1989maintenance, mcauley1990regulation, stabell2002ecological, iwai2015maintenance}. We note that, in contrast to \cite{iwai2015maintenance}, for example, our model highlights the importance of horizontal transmission in determining the stable symbiont distribution. In fact horizontal transmission of symbionts is necessary, when combined with host control, in order to react appropriately to environmental changes (cf. \cite{lowe2015shining}). 

The model presented here forms an excellent basis for further study of endosymbiotic relationships. Our model gives a number of detailed predictions regarding the photosymbiotic relationship between \emph{P. bursaria} and \emph{Chlorella}, in particular, the dependence of symbiont load on light and host control. These predictions provide insight into the underlying mechanisms and trading relationships between the two partners. Subsequent investigations into these details may permit greater understanding of the nature of the relationship between the two partners including its equality, its history, its future trajectory and ultimately greater understanding of this important route to complex lifeforms.

\section*{Acknowledgements}
The authors are grateful to the editor and referees for their insightful comments, leading to substantial improvements to the original manuscript. We are also grateful to Richard Law for helpful discussions throughout the development of this work, and to NERC for funding this research.




\bibliographystyle{elsarticle-num} 
\bibliography{Refs}





\end{document}